\documentclass[reprint,aps]{revtex4-1}
\usepackage{amsmath}
\usepackage{amssymb}
\usepackage{graphicx}

\begin{document}
\title{Unitary dynamics and finite-time measurements: a case study}
\author {Andreas Prinz-Zwick}
\author{Gert-Ludwig Ingold}
\author{Peter Talkner}
\affiliation{Institut f\"ur Physik, Universit\"at Augsburg,
             Universit\"atstra{\ss}e 1, D-86135 Augsburg, Germany}

\begin{abstract}
The inhibition of the decay of a quantum system by frequent measurements is
known as quantum Zeno effect. Beyond the limit of projective measurements, the
interplay between the unitary dynamics of the system and the coupling to a
measurement apparatus becomes relevant. We explore this interplay by
considering a quantum particle moving on a one-dimensional chain. A local
measurement by coupling to an apparatus with a two-dimensional Hilbert space
detects the presence of the particle on a specific chain site. The decay of the
population is studied analytically for a two-site chain and numerically for a
larger system as a function of the measurement time and the time between
subsequent measurements.  Particular attention is given to the shift of the
energy of the measured site due to the coupling to the apparatus. The decay of
the initial population can be hindered or accelerated, depending on the chosen
system and the coupling parameters.
\end{abstract}
\maketitle

\section{Introduction}

Quantum mechanics does not allow the observation of a system without perturbing
it. This important difference to classical mechanics nowadays has lead to
applications, e.g.\ in the detection of an eavesdropper in quantum
communication \cite{gisin02}. Another spectacular effect known for almost forty
years is the quantum Zeno effect \cite{misra77,chiu77}. Here, measurements of a
system which are repeated in short time intervals can lead to a slowing-down of
its dynamics which can even come to a halt when the time between subsequent
measurements goes to zero.

Even though the first experiment performed on beryllium ions was initially
analysed in terms of projective measurements \cite{itano90}, a more appropriate
description in terms of the corresponding Bloch equations shows that the
experimental results can well be understood without assuming a projective
measurement \cite{block91,frerichs91}. In fact, while many studies of the
quantum Zeno effect assume a projective measurement, a treatment of the
measurement in terms of a coupling to some sort of measurement apparatus is
certainly more realistic. For a recent review of a dynamical description of
quantum measurements, we refer to ref.~\cite{allahverdyan13}. Generalised
measurements in the context of the quantum Zeno effect and their experimental
realisation are discussed in \cite{mack14}.

The quantum Zeno effect with a non-projective measurement was studied for the
decay of a driven two-level system coupled to an electromagnetic field
\cite{ruseckas01a,ruseckas01b}. Instead of coupling the measurement device
directly to the system to be measured, also indirect measurement processes have
been explored where decay products are detected \cite{shaji04,koshino05}.

In the present paper, we are interested in the dynamics of a particle on 
a finite one-dimensional chain where the presence or absence of the particle
at a selected site is measured by coupling to a measurement apparatus. As
no decay products are generated when the particle leaves the selected site,
we necessarily need to perform a direct measurement of its position. 

Such a situation was first studied by Gurvitz \cite{gurvitz00} who was
interested in the delocalisation induced by a local measurement on a
one-dimensional disordered system described by an Anderson model. There, the
transmission through a quantum point contact was used to continuously monitor
the presence of an electron in one of the quantum dots of an array of coupled
dots. It was found that even for very weak coupling between the electron on the
quantum dot and the electrons flowing through the quantum point contact,
delocalisation on the quantum dot array would ensue.

The asymptotic state, and the approach to it, of a particle moving on a chain
of sites under the influence of repeated projective measurements were
investigated in several works \cite{flores99,gordon10,yi11}.  Here, we
generalise the study presented in ref.~\onlinecite{yi11} to the case of
finite-time measurements and discuss the dependence on the measurement time.

\begin{figure}
 \includegraphics[width=\columnwidth]{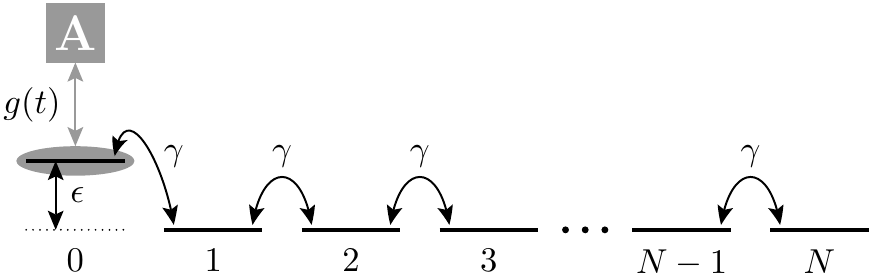}
 \caption{The setup considered here consists of a system represented by a
  chain of $N$ sites between which the particle can hop with an amplitude
  $\gamma$. The energy of the left-most site can be shifted by an amount
  $\epsilon$. The presence of the particle on this site is measured by
  coupling an apparatus A to the site with a coupling constant $g$.}
 \label{fig:setup}
\end{figure}

To this end, we consider the setup displayed in figure~\ref{fig:setup} where a
particle can move along a one-dimensional chain containing $N+1$ sites. The
presence of the particle on the left-most site of the chain indicated as
site~0, is measured by an apparatus A which is coupled to this site during the
measurement periods.  The apparatus, its initial state and the coupling to the
system to be measured is chosen according to a model proposed by Zurek
\cite{zurek00} and will be described in more detail in
section~\ref{sec:apparatus_model}.

According to the previous description, the Hamiltonian
\begin{equation}
\hat H = \hat H_\text{c}\otimes\openone_\text{A}+g(t)\hat H_\text{I}
\end{equation}
consists of a part $\hat H_\text{c}$ describing the motion on the chain while
$\hat H_\text{I}$ pertains to the coupling to the measurement apparatus. The
unperturbed apparatus is supposed to be static with a vanishing apparatus
Hamiltonian $H_\text{A}$.  The time dependence of the coupling constant $g(t)$
ensures that the coupling is present only during measurement periods while it
vanishes between them when the particle moves unobserved on the chain. The
identity operator in the Hilbert space of the apparatus is denoted by
$\openone_\text{A}$. The chain Hamiltonian is given by
\begin{equation}
\label{eq:hamiltonian_chain}
\hat H_\text{c} = \epsilon\vert 0\rangle\langle 0\vert
-\gamma\sum_{n=0}^{N-1}\left(\vert n\rangle\langle n+1\vert
+\vert n+1\rangle\langle n\vert\right)\,.
\end{equation}
Here, the hopping matrix element is denoted by $\gamma$. The left-most site is
shifted in energy relative to the other sites by an amount $\epsilon$. 
The explicit form of the Hamiltonian $\hat H_\text{I}$ will be given below in
(\ref{eq:interaction_hamiltonian_discrete}).

The measurement apparatus and the particle on the chain are supposed to be
initially in a factorised state. During the measurement process, correlations
are built up resulting in an entangled state. After decoupling the apparatus,
the system is thus typically left in a mixed state. We will not read out the
state of the apparatus and thus only carry out a premeasurement which still
allows for all possible measurement outcomes. The apparatus will be reset to
its initial state before it is coupled again to the left-most site.

We start by reviewing Zurek's model in section~\ref{sec:apparatus_model} and
adapting it to our purpose. In section~\ref{sec:minimal_model} we construct a
minimal model where the length of the chain consists of just two lattice sites.
This minimal model allows us to study the interplay between the measurement
dynamics and the system dynamics during the finite measurement time. We pay
particular attention to an energy shift of the measured site due to the
coupling to the measurement apparatus. Section~\ref{sec:repeated_measurements}
is devoted to the discussion of numerical results for a chain of more than two
sites and the difference of the dynamics to the case of projective
measurements. Finally, in section~\ref{sec:conclusions} we present our
conclusions.

\section{Model for the measurement apparatus}
\label{sec:apparatus_model}

In this section we restrict ourselves to the measurement aspect of the setup
displayed in figure~\ref{fig:setup} and neglect any proper dynamics of the
system during the time in which the measurement apparatus is interacting with
the system. We thus disregard the chain Hamiltonian $\hat H_\text{c}$ or at
least assume that the system dynamics is much slower than the dynamics due to
the coupling between the system and the measurement apparatus. The interplay
between system dynamics and measurement process will be the subject of
section~\ref{sec:minimal_model}.

As a motivation for the model describing the measurement apparatus and its
coupling to the chain to be used later on, we follow von Neumann
\cite{vonNeumann55} and consider an apparatus with a pointer described by the
continuous position operator $\hat x$. The initial position of the pointer is
assumed to be at the origin, i.e.\ its initial state is given by $\vert
0\rangle_\text{A}$. Here, $\vert 0\rangle_\text{A}$ is the eigenstate of the
operator of the pointer position $\hat x$ with eigenvalue 0, i.e.\ $\hat x
\vert 0\rangle_\text{A}=0$. Assuming that the system to be measured is in an
eigenstate $\vert s\rangle$ of the observable
\begin{equation}
\hat s = \int\text{d}s' s'\vert s'\rangle\langle s'\vert,
\end{equation}
the pointer after completion of the measurement process should be found at the
position $x=s$, i.e. in the eigenstate $\vert s\rangle_\text{A}$ of the
position operator $\hat x$ of the apparatus. Such a transition can be achieved
by the translation operator
\begin{equation}
\hat T = \exp\left(-\frac{\text{i}}{\hbar}s\hat p\right)\,,
\end{equation}
where the momentum operator $\hat p$ conjugate to $\hat x$ operates on the
Hilbert space of the apparatus.

For a general initial state of the system, the coupling between system and
apparatus can then be described by an interaction Hamiltonian $g(t)\hat H_\text{I}$
with
\begin{equation}
\label{eq:interaction_hamiltonian_continuous}
\hat H_\text{I} = \hat s\hat p\,.
\end{equation}
The time-dependent coupling constant $g(t)$ ensures that the coupling is only
present during the measurement process. By choosing its magnitude, we can
control how strongly the apparatus is coupled to the system. During a
measurement, when $g$ takes on a fixed value, the translation of the pointer
described above can then be achieved by the time evolution operator
\begin{equation}
\label{eq:time_evolution_continuous}
\hat U(t) = \exp\left(-\frac{\text{i}}{\hbar}g\hat s\hat p t\right)
\end{equation}
if we choose the measurement time as $t_\text{m} = 1/g$.

For an arbitrary initial state of the system described by the function $c(s')$
and the apparatus in its initial state $\vert 0\rangle_\text{A}$, the time
evolution of system and apparatus starts from the state
\begin{equation}
\vert\Psi(0)\rangle = \int\text{d}s' c(s')\vert s'\rangle\vert0\rangle_\text{A}\,.
\end{equation}
At the end of the measurement process described by the time evolution operator
(\ref{eq:time_evolution_continuous}) we obtain the entangled state
\begin{equation}
\vert\Psi(1/g)\rangle = \int\text{d}s'c(s')\vert s'\rangle\vert s'\rangle_\text{A}\,.
\end{equation}
The value of the measured system observable $\hat s$ is now encoded in the 
pointer state of the apparatus.

For our purposes, it is sufficient to provide the apparatus with a finite
Hilbert space. Since we only need to measure whether the first lattice site is
occupied or not, we will eventually restrict the Hilbert space to two
dimensions.  For the moment, however, we assume the Hilbert space of the
apparatus to be $N$-dimensional.

From the discussion above it turns out as useful to introduce, apart from the
pointer states $\vert x\rangle_\text{A}$, eigenstates $\vert p\rangle_\text{A}$
of the momentum operator $\hat p$ conjugate to the pointer position operator
$\hat x$.  Accordingly, in the discrete case, we introduce states $\{\vert
A_j\rangle_\text{A}\}$ and $\{\vert B_k\rangle_\text{A}\}$ which correspond to
the pointer states and the complementary momentum basis, respectively. The two
sets of states are related by
\begin{equation}
\vert A_j\rangle_\text{A} = \frac{1}{N^{1/2}}\sum_{k=0}^{N-1}
    \exp\left(-\frac{2\pi\text{i}}{N}jk\right)\vert B_k\rangle_\text{A}
\end{equation}
Following ref.~\cite{zurek00}, we define a discrete version of the interaction
Hamiltonian between system and apparatus
\begin{equation}
\label{eq:interaction_hamiltonian_discrete}
\hat H_\text{I} = \hat s\hat B
\end{equation}
with the system observable
\begin{equation}
\hat s = \sum_{k=0}^{N-1} k\vert s_k\rangle\langle s_k\vert
\end{equation}
and the apparatus operator
\begin{equation}
\label{eq:apparatus_operator_discrete}
\hat B = \sum_{k=0}^{N-1} k\vert B_k\rangle_\text{A}{}_\text{A}\langle B_k\vert\,. 
\end{equation}
For a measurement time
\begin{equation}
\label{eq:measurement_time_discrete}
t_\text{m} = \frac{2\pi\hbar}{gN}
\end{equation}
an initial state $\vert s_j\rangle\vert A_0\rangle_\text{A}$ evolves into
\begin{equation}
\label{eq:measurement_result}
\exp\left(-\frac{\text{i}}{\hbar}g\hat H_\text{I}t_\text{m}\right)\vert s_j\rangle
\vert A_0\rangle_\text{A} = \vert s_j\rangle\vert A_j\rangle_\text{A}
\end{equation}
by the action of the interaction Hamiltonian
(\ref{eq:interaction_hamiltonian_discrete}), as required for a perfect
measurement. A general initial pure state of the system will thus be turned
into an entangled state between system and measurement apparatus
\begin{equation}
\exp\left(-\frac{\text{i}}{\hbar}g\hat H_\text{I}t_\text{m}\right)
\sum_j c_j\vert s_j\rangle \vert A_0\rangle_\text{A} =
\sum_j c_j\vert s_j\rangle\vert A_j\rangle_\text{A}\,.
\end{equation}

We are now in the position to specialise the model just described to the case
$N=2$. However, we also need to slightly generalise it for our purposes. As we
shall see, the coupling (\ref{eq:interaction_hamiltonian_discrete}) of the
chain to the measurement apparatus effectively changes the on-site energy at
the left-most lattice site. This effect is reminiscent of the Lamb shift due to
the coupling of a bound electron to the electromagnetic field \cite{bethe47} or
the potential renormalisation of a dissipative system coupled to its
environment \cite{weiss12}.  A change of the on-site energy on one lattice site
will necessarily influence the dynamics on the chain. In addition, as an effect
of the coupling to the measurement apparatus, it will be time-dependent. During
the measurement process, the on-site energy will be modified while it assumes
its bare value $\epsilon$ between measurements.

For the discussion of the Zeno effect on a chain, we make use of the
interaction Hamiltonian (\ref{eq:interaction_hamiltonian_discrete}). The
system observable that detects the presence of the particle on the left-most
lattice site is given by
\begin{equation}
\hat s = \vert 0\rangle\langle 0\vert\,.
\end{equation}

The measurement operator is obtained from
(\ref{eq:apparatus_operator_discrete}) by setting $N=2$. In addition, for later
convenience, we want to be able to shift the spectrum in order to analyse the
effect of the coupling-induced shift of the energy of the measured site. We
thus define
\begin{equation}
\label{eq:apparatus_operator_2}
\hat B = \delta\openone_\text{A}
         -\frac{1}{2}\vert B_0\rangle_\text{A}{}_\text{A}\langle B_0\vert
         +\frac{1}{2}\vert B_1\rangle_\text{A}{}_\text{A}\langle B_1\vert\,,
\end{equation}
where the parameter $\delta$ allows to shift the spectrum of the operator $\hat
B$.  For $\delta=1/2$, we recover the definition
(\ref{eq:apparatus_operator_discrete}) for $N=2$.

For $N=2$, the states $\{\vert A_0\rangle_\text{A}, \vert
A_1\rangle_\text{A}\}$ and $\{\vert B_0\rangle_\text{A}, \vert
B_1\rangle_\text{A}\}$ are related by a Hadamard transform
\begin{equation}
\begin{pmatrix} \vert B_0\rangle_\text{A} \\ \vert B_1\rangle_\text{A} \end{pmatrix} =
\frac{1}{2^{1/2}}\begin{pmatrix} 1 & 1 \\ 1 & -1 \end{pmatrix}
\begin{pmatrix} \vert A_0\rangle_\text{A} \\ \vert A_1\rangle_\text{A} \end{pmatrix}\,.
\end{equation}
Note that the Hadamard matrix equals its own inverse so that the 
back transformation is of the same form. Finally, the measurement time
(\ref{eq:measurement_time_discrete}) becomes
\begin{equation}
\label{eq:measurement_time_2}
t_\text{m} = \frac{\pi\hbar}{g}\,.
\end{equation}
In the limit $g\to\infty$, we obtain $t_\text{m}=0$. It should be expected that then
the limit of a projective measurement is recovered. We will check this explicitly
for a two-site chain in the next section.

\section{Competition between system dynamics and measurement dynamics}
\label{sec:minimal_model}

The model discussed in this paper contains two essential ingredients. The
measurement takes a finite time during which the system is coupled to the
measurement apparatus. This dynamics competes with the system's own dynamics on
the finite chain.

To gain insight into the interplay between these two dynamical mechanisms, we
consider a minimal model where the chain consists of two sites, one of them
being measured by an apparatus with a two-dimensional Hilbert space. The
corresponding Hamiltonian governing the dynamics of the combined system
consisting of chain and apparatus reads
\begin{equation}
\label{eq:hamiltonian_minimal}
\begin{aligned}
\hat H &= \left[\epsilon\vert0\rangle\langle0\vert-\gamma\big(\vert0\rangle\langle1\vert
+\vert1\rangle\langle0\vert\big)\right]\otimes\openone_\text{A}\\
&\quad+g(t)\vert0\rangle\langle0\vert\otimes\bigg(\delta\openone_\text{A}
-\frac{1}{2}\vert B_0\rangle_\text{A}{}_\text{A}\langle B_0\vert\\
&\hspace{3truecm}+\frac{1}{2}\vert B_1\rangle_\text{A}{}_\text{A}\langle B_1\vert\bigg)\,,
\end{aligned}
\end{equation}
where during a measurement the coupling constant $g(t)$ takes a constant
non-zero value $g$ while it vanishes during periods of free dynamics.

The bare on-site energy $\epsilon$ of the left-most site is modified by the
coupling to the apparatus and becomes $\epsilon+g\delta$ during the measurement
processes.  We will concentrate on the special case $\epsilon=0$ and, at the
end of this section, comment on how the results are modified for non-zero
values of $\epsilon$.

The Hamiltonian (\ref{eq:hamiltonian_minimal}) with $\epsilon=0$ can be easily
diagonalised yielding the eigenenergies
\begin{equation}
\label{eq:e0pm}
E_{0\pm} = \frac{g}{4}(2\delta-1)\pm\left(\gamma^2+\frac{g^2}{16}(2\delta-1)^2\right)^{1/2}
\end{equation}
and 
\begin{equation}
\label{eq:e1pm}
E_{1\pm} = \frac{g}{4}(2\delta+1)\pm\left(\gamma^2+\frac{g^2}{16}(2\delta+1)^2\right)^{1/2}.
\end{equation}
With
\begin{equation}
\tan(\phi_0) = -\frac{4\gamma}{g(2\delta-1)}
\end{equation}
and
\begin{equation}
\tan(\phi_1) = -\frac{4\gamma}{g(2\delta+1)}
\end{equation}
where $0\le\phi_0, \phi_1<\pi$, the corresponding eigenstates can be expressed as
\begin{equation}
\begin{aligned}
\vert 0+\rangle &= \cos\left(\frac{\phi_0}{2}\right)\vert 0\rangle\vert B_0\rangle_\text{A}
                    +\sin\left(\frac{\phi_0}{2}\right)\vert 1\rangle\vert B_0\rangle_\text{A}\\
\vert 0-\rangle &= \sin\left(\frac{\phi_0}{2}\right)\vert 0\rangle\vert B_0\rangle_\text{A}
                    -\cos\left(\frac{\phi_0}{2}\right)\vert 1\rangle\vert B_0\rangle_\text{A}
\end{aligned}
\end{equation}
and
\begin{equation}
\begin{aligned}
\vert 1+\rangle &= \cos\left(\frac{\phi_1}{2}\right)\vert 0\rangle\vert B_1\rangle_\text{A}
                    +\sin\left(\frac{\phi_1}{2}\right)\vert 1\rangle\vert B_1\rangle_\text{A}\\
\vert 1-\rangle &= \sin\left(\frac{\phi_1}{2}\right)\vert 0\rangle\vert B_1\rangle_\text{A}
                    -\cos\left(\frac{\phi_1}{2}\right)\vert 1\rangle\vert B_1\rangle_\text{A}\,.
\end{aligned}
\end{equation}

We first check whether the limit of infinite coupling, $g\to\infty$, leads to a
projective measurement as expected in the previous section. As the initial state,
we choose
\begin{equation}
\label{eq:initial_state_2}
\vert\Psi(0)\rangle = (c_0\vert 0\rangle+c_1\vert 1\rangle)\vert A_0\rangle_\text{A}
\end{equation}
with complex coefficients $c_0$ and $c_1$ obeying the normalisation condition
$\vert c_0\vert^2+\vert c_1\vert^2=1$. The evolution of this state is subject
to the Hamiltonian (\ref{eq:hamiltonian_minimal}). The result of a premeasurement
is the reduced density matrix of the chain
\begin{equation}
\label{eq:system_density_matrix}
\rho^\text{S}(t_\text{m}) = \text{Tr}_\text{A}\big(\vert\Psi(t_\text{m})\rangle
\langle\Psi(t_\text{m})\vert\big)
\end{equation}
at the measurement time defined in (\ref{eq:measurement_time_2}). Here,
$\text{Tr}_\text{A}$ denotes the trace over the Hilbert space of the apparatus.
This density matrix has to be compared with the result
\begin{equation}
\rho_\text{proj} = \begin{pmatrix}\vert c_0\vert^2 & 0\\0 & \vert c_1\vert^2\end{pmatrix}
\end{equation}
of a projective measurement on the chain state $c_0\vert 0\rangle+c_1\vert1\rangle$.

To quantify the difference between the reduced density matrix of the chain after
the premeasurement $\rho^\text{S}(t_\text{m})$ and $\rho_\text{proj}$, we determine
the trace distance
\begin{equation}
\label{eq:tracedistance}
T(\rho^\text{S}(t_\text{m}), \rho_\text{proj}) = \frac{1}{2}\text{Tr}_\text{S}
\left(\vert\rho^\text{S}(t_\text{m})-\rho_\text{proj}\vert\right)\,,
\end{equation}
where $\text{Tr}_\text{S}$ denotes the trace over the system Hilbert space. In
the limit of large values of the coupling between system and measurement
apparatus, the trace distance becomes proportional to $1/g$ or, equivalently,
proportional to the measurement time $t_\text{m}$, i.e.,
\begin{equation}
\label{eq:tracedistance_lin}
T(\rho^\text{S}(t_\text{m}), \rho_\text{proj}) = \frac{2\gamma}{g}T_1+O(1/g^2)\,.
\end{equation}
For a general initial state of the form (\ref{eq:initial_state_2}) one finds
\begin{equation}
\label{eq:tracedistance1}
T_1 = \left\vert\frac{2\delta-\sin(\pi\delta)}{1-4\delta^2}(c_0^2+c_1^2)
-\text{i}\frac{\cos(\pi\delta)}{1-4\delta^2}(c_0^2-c_1^2)\right\vert\,.
\end{equation}
For $\delta=0$ and an initial state (\ref{eq:initial_state_2}) with
$c_0/c_1=\pm1$, the term of order $1/g$ vanishes and the projective limit is
approached as $1/g^2$ for large $g$.

\begin{figure}
 \includegraphics[width=\columnwidth]{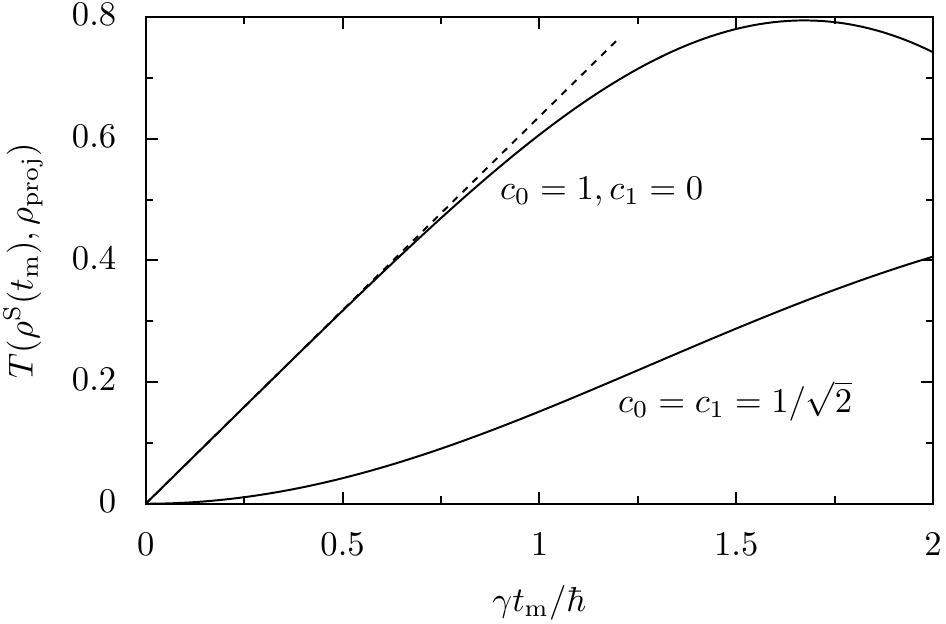}
 \caption{The trace distance (\ref{eq:tracedistance}) between reduced density
  matrices $\rho^\text{S}(t_\text{m})$ and $\rho_\text{proj}$ corresponding to a
  measurement by coupling to an apparatus and a projective measurement, respectively,
  is displayed as a function of the duration $t_\text{m}$ of the measurement for
  $\delta=0$ and $\gamma=1$. The upper curve corresponds to an initial state of the form
  (\ref{eq:initial_state_2}) with $c_0=1$ and $c_1=0$. The dashed line represents
  its linear approximation given by (\ref{eq:tracedistance_lin}) with
  (\ref{eq:tracedistance1}). The lower curve with $c_0=c_1=2^{-1/2}$ represents a
  special case where the linear term vanishes.}
\label{fig:tracedistance}
\end{figure}

Figure~\ref{fig:tracedistance} displays the trace distance
(\ref{eq:tracedistance}) as a function of the measurement time
$t_\text{m}=\pi\hbar/g$ for $\delta=0$. The upper curve corresponds to an initial
state where the system is localised on the left lattice site. Its linear approximation
according to (\ref{eq:tracedistance_lin}) with (\ref{eq:tracedistance1}) is depicted
as dashed line. The lower curve with $c_0=c_1=2^{-1/2}$ represents a special case
where the linear approximation vanishes and the projective limit is approached much
faster than in the general case.

\begin{figure}
 \includegraphics[width=\columnwidth]{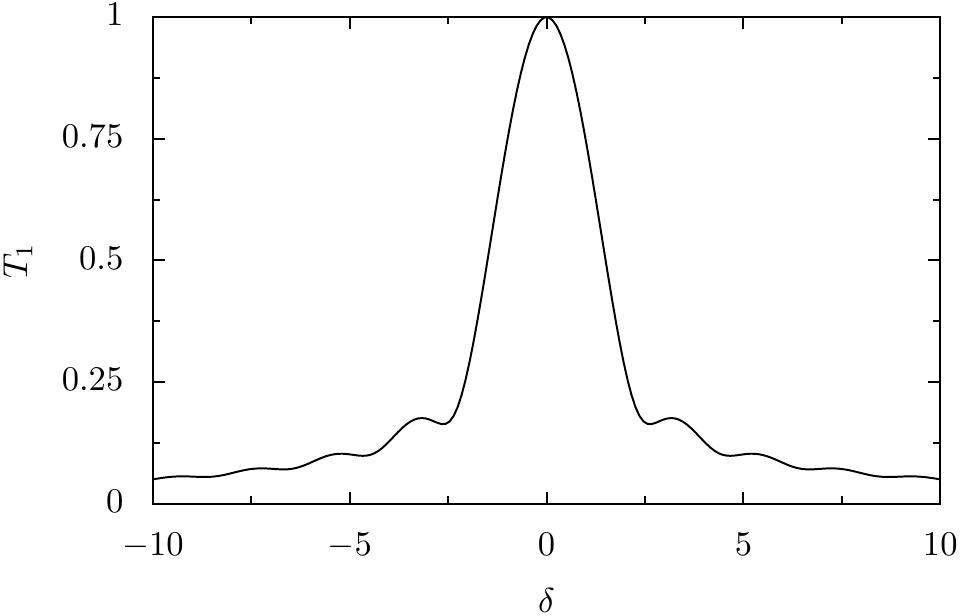}
 \caption{The coefficient $T_1$ defined in (\ref{eq:tracedistance1})
  as a function of the parameter $\delta$ for an initial state of the form
  (\ref{eq:initial_state_2}) with $c_0=1, c_1=0$.}
\label{fig:t1_delta}
\end{figure}

The dependence of the coefficient $T_1$ appearing in the linear approximation
of the trace distance (\ref{eq:tracedistance}) on $\delta$ is shown in
figure~\ref{fig:t1_delta} for an initial state of the form
(\ref{eq:initial_state_2}) with $c_0=1$ and $c_1=0$. The maximum at $\delta=0$
can be explained in terms of the system dynamics during the measurement.  A
non-zero value of $\delta$ effectively shifts the on-site energy of the left
site out of resonance with the right site, thus suppressing the system
dynamics. As a consequence $T_1$ decreases with increasing absolute values of
$\delta$. For degenerate on-site energies on the chain, approaching the
projective limit requires a particularly short measurement time $t_\text{m}$
or, equivalently, strong coupling between chain and measurement apparatus.

The influence of the system dynamics on a finite-time measurement
implies that the resulting state of system and apparatus is not given by
(\ref{eq:measurement_result}) where the system dynamics was neglected
completely.  For the initial state
\begin{equation}
\vert\Psi(0)\rangle = \vert 0\rangle\vert A_0\rangle_\text{A}
\end{equation}
one finds for the state at the end of the measurement process to leading order
in the inverse coupling constant
\begin{equation}
\label{eq:state_system_apparatus}
\begin{aligned}
\vert\Psi(t_\text{m})\rangle &=
\text{ie}^{-\text{i}\pi\delta}\vert0\rangle\vert A_1\rangle_\text{A}\\
&\quad+\frac{\gamma}{g}\left(\frac{1-\text{ie}^{-\text{i}\pi\delta}}{2\delta-1}+
                       \frac{1+\text{ie}^{-\text{i}\pi\delta}}{2\delta+1}\right)
                 \vert 1\rangle\vert A_0\rangle_\text{A}\\
&\quad+\frac{\gamma}{g}\left(\frac{1-\text{ie}^{-\text{i}\pi\delta}}{2\delta-1}-
                       \frac{1+\text{ie}^{-\text{i}\pi\delta}}{2\delta+1}\right)
                 \vert 1\rangle\vert A_1\rangle_\text{A}\\
&\quad+O(1/g^2)\,.
\end{aligned}
\end{equation}
The first term corresponds to a measurement that is not influenced by the system
dynamics.  For $\delta=1/2$, it agrees with the expected state $\vert
0\rangle\vert A_1\rangle_\text{A}$, where the state $\vert A_1\rangle_\text{A}$
indicates the presence of the particle on the measured site 0. A contribution
with the particle on site 0 and the apparatus in state $\vert
A_0\rangle_\text{A}$ is only found if terms of order $1/g^2$ are retained.

The influence of the system dynamics on the state resulting from a measurement
can again be quantified by means of the trace distance. To leading order in
$1/g$, the trace distance between the density matrix
$\rho(t_\text{m})=\vert\Psi(t_\text{m})\rangle \langle\Psi(t_\text{m})\vert$
and the density matrix $\rho_0 = \vert 0\rangle \langle 0\vert\otimes\vert
A_1\rangle_\text{A}{}_\text{A}\langle A_1\vert$ is found as
\begin{equation}
T(\rho(t_\text{m}), \rho_0) = \frac{2^{3/2}\gamma}{g}T_1'+O(1/g^2)
\end{equation}
with 
\begin{equation}
T_1' = \left[\frac{1-\sin(\pi\delta)}{2(2\delta-1)^2}
            +\frac{1+\sin(\pi\delta)}{2(2\delta+1)^2}\right]^{1/2}\,.
\end{equation}
The latter expression agrees with (\ref{eq:tracedistance1}) for $c_0=1, c_1=0$.
The dependence on the parameter $\delta$ of the perturbation of the measurement
result by the system dynamics can thus again be inferred from
figure~\ref{fig:t1_delta} and the corresponding discussion given above.

After having explored how the system dynamics perturbs the measurement process,
we now address the question of how the system dynamics is affected by the
coupling to the measurement apparatus. To this end we consider the probability
to find the particle on site~0
\begin{equation}
\rho_{00}^\text{S}(t) = \langle0\vert\rho^\text{S}(t)\vert0\rangle\,,
\end{equation}
where the reduced density matrix of the system has been defined in
(\ref{eq:system_density_matrix}).

Starting from the pure initial state
\begin{equation}
\vert\Psi(0)\rangle = \vert0\rangle\vert A_0\rangle_\text{A}\,,
\end{equation}
diagonalisation of the Hamiltonian (\ref{eq:hamiltonian_minimal}) yields
\begin{equation}
\rho^S_{00}(t) = 1 + \frac{\gamma^2}{\hbar^2}\left[\frac{\cos(\Omega_0 t)-1}{\Omega_0^2}
                   + \frac{\cos(\Omega_1 t)-1}{\Omega_1^2}\right]\,.
\end{equation}
Here,
\begin{equation}
\Omega_n = \frac{E_{n+}-E_{n-}}{\hbar} \quad\text{for $n=0,1$}
\end{equation}
are differences of the energies defined in (\ref{eq:e0pm}) and (\ref{eq:e1pm}).
Of particular interest is the decay of the population on the left site of the
chain which for short times is given by
\begin{equation}
\label{eq:taylor}
\rho_{00}^\text{S}(t) = 1-\frac{\gamma^2t^2}{\hbar^2}
+\frac{1}{3}\left(1+\frac{g^2(4\delta^2+1)}{16\gamma^2}\right)
\frac{\gamma^4t^4}{\hbar^4}+O(t^6)\,.
\end{equation}
The leading decay term of second order in $t$ is exclusively due to the
hopping between the two sites of the chain. The coupling to the measurement
apparatus appears first in fourth order in $t$ and tends to delay the
decay of the population on the measured site. Including a non-zero value for
$\delta$ in the coupling part of (\ref{eq:hamiltonian_minimal}) helps to suppress
further the dynamics on the chain.

\begin{figure}
 \includegraphics[width=\columnwidth]{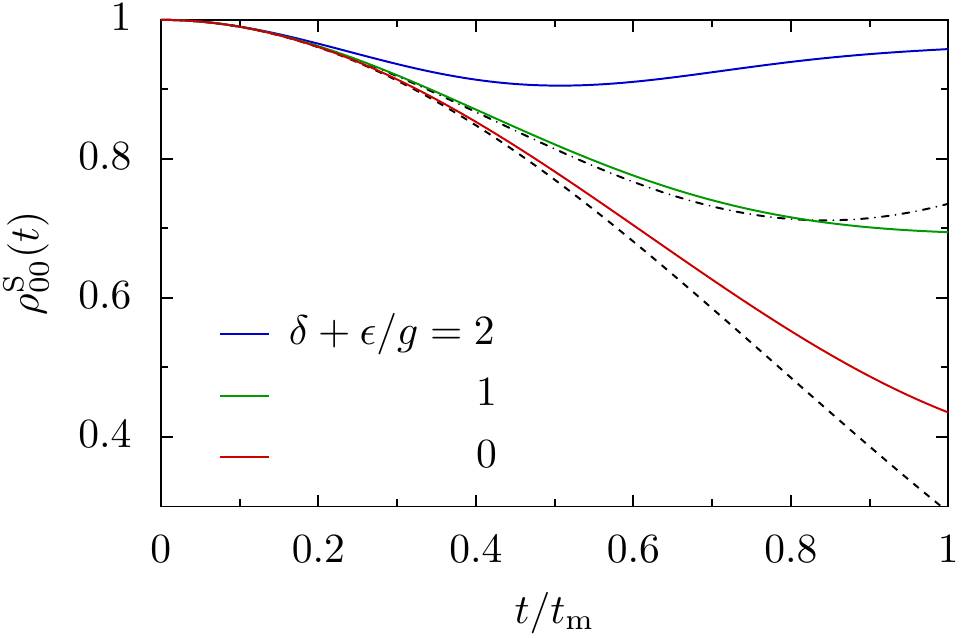}
 \caption{The survival probability on the left site of the chain during a
  measurement with $g/\gamma=\pi$ is shown for $\delta+\epsilon/g = 0, 1$, and
  $2$ increasing from the lower to the upper solid line. The free evolution
  in the absence of an apparatus, i.e.\ $g=0$, is depicted as dashed line for
  degenerate on-site energies and as dash-dotted line if the left site is
  lowered by $\epsilon/\gamma=-\pi$ with respect to the right site.}
 \label{fig:k0min}
\end{figure}

In figure~\ref{fig:k0min} the survival probability is presented during a
measurement with coupling constant $g/\gamma=\pi$. Clearly, increasing $\delta$
tends to hinder the dynamics on the chain in agreement with the fourth-order
term in (\ref{eq:taylor}). Compared to the free dynamics on the chain,
$g/\gamma=0$, depicted as dashed line, this effect is visible for all values of
$\delta$. Only at the beginning of the measurement process, the decay is
independent of the coupling to the measurement apparatus as expressed by the
second-order term in (\ref{eq:taylor}).

However, the energy shift induced by the coupling to the apparatus can also
enhance the decay of the initial chain state. This is the case if the two
on-site energies are not degenerate, i.e.\ if the energy shift $\epsilon$ in
the chain part of the Hamiltonian (\ref{eq:hamiltonian_minimal}) is non-zero.
As we have discussed when introducing that Hamiltonian, the energy of site~0
under the influence of the coupling to the apparatus is given by
$\epsilon+g\delta$.  The fastest decay of the initial state is then no longer
obtained for $\delta=0$ but for $\delta=-\epsilon/g$. In
figure~\ref{fig:k0min}, the dash-dotted line depicts the time dependence of
$\rho_{00}^\text{S}(t)$ in the absence of a coupling to the apparatus and for
an energy shift of $\epsilon/\gamma=-\pi$.  For $\delta=1$, corresponding to
$\delta+\epsilon/g=0$ in the figure, the decay of the initial state is
optimally accelerated.

\section{Repeated measurements}
\label{sec:repeated_measurements}

So far, we have focussed on the interplay between the system dynamics and the
coupling to the apparatus during a single measurement process. Now, we will
consider repeated measurements. At the beginning of the $n$-th measurement, the
particle on the chain will generally be in a mixed state described by the
density matrix $\rho_n^\text{S}$. The initial state of the apparatus will
always be given by the pure state $\vert A_0\rangle_\text{A}$ as explained in
section~\ref{sec:apparatus_model}. During the $n$-th measurement process, an
entangled state between system and apparatus will evolve which at the end of
the measurement is described by a density matrix $\rho_n$. We will not read out
the measurement result and thus account for all sequences of measurement
outcomes during the repeated measurements. 

After the $n$-th measurement, the free system dynamics starts with an initial
density matrix $\text{Tr}_\text{A}(\rho_n)$ where the degrees of freedom of the
apparatus have been traced out. At the end of the free time evolution, the
system state is given by the density matrix $\rho_{n+1}^\text{S}$. One cycle
spanning the time between the beginning of subsequent measurement processes can
then be described by the sequence of density matrices
\begin{equation}
\begin{aligned}
\rho_n^\text{S}\otimes\vert A_0\rangle_\text{A}{}_\text{A}\langle A_0\vert
\rightarrow \rho_n
\rightarrow \text{Tr}_\text{A}(\rho_n)
\rightarrow \rho_{n+1}^\text{S}\\
\rightarrow \rho_{n+1}^\text{S}\otimes\vert A_0\rangle_\text{A}{}_\text{A}\langle A_0\vert
\end{aligned}
\end{equation}

\begin{figure}
 \includegraphics[width=\columnwidth]{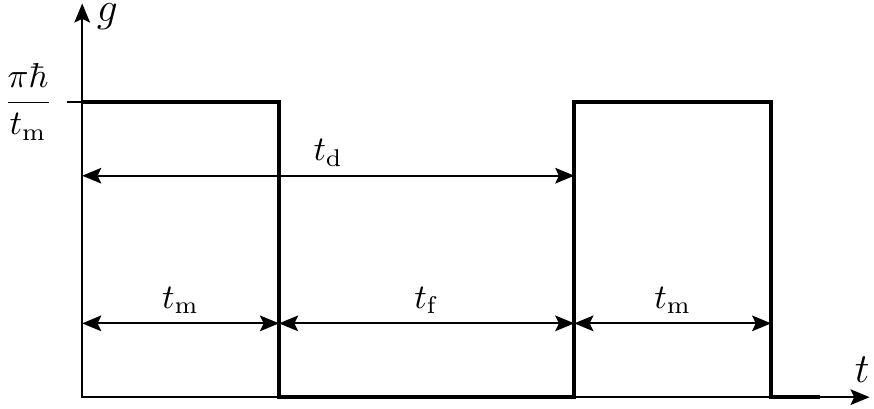}
 \caption{Time scales involved in repeated measurements: Measurements of duration
  $t_\text{m}$ are separated by periods of duration $t_\text{f}$ of free evolution.
  The time between the beginning of subsequent measurements is $t_\text{d}=
  t_\text{m}+t_\text{f}$. The coupling strength between system and apparatus
  satisfies (\ref{eq:measurement_time_2}).}
 \label{fig:timescales}
\end{figure}

The relevant times involved in such a cycle are depicted in
figure~\ref{fig:timescales}.  The duration $t_\text{m}$ of a single measurement
is related to the coupling constant $g$ between system and apparatus by
(\ref{eq:measurement_time_2}). After the measurement, a period of length
$t_\text{f}$ follows where the system is decoupled from the apparatus and
evolves freely. The lapse of time between the beginning of subsequent
measurements is then given by $t_\text{d}=t_\text{m}+t_\text{f}$. For the
following discussion it should be kept in mind that obviously the time span
$t_\text{d}$ can never be smaller than the duration of a measurement
$t_\text{m}$.

The results for the probability $\rho_{00}^\text{S}$ to find the particle on
the left-most chain site presented in the figures of this section have been
calculated numerically for a chain of 15 sites. For $\epsilon/\gamma=0$, the
maximum group velocity on the chain is given by $2\gamma/\hbar$, so that in the
absence of any measurements a particle initially located on site~0 can be
expected to reappear there for the first time when $\gamma t/\hbar\sim 15$.
This estimation is found to be correct even in the presence of projective
measurements \cite{yi11} and can be expected to hold also for finite-time
measurements. In order to avoid effects arising from reflections at the right
end of the chain, we will choose times smaller than $15\hbar/\gamma$.

\begin{figure}
 \includegraphics[width=\columnwidth]{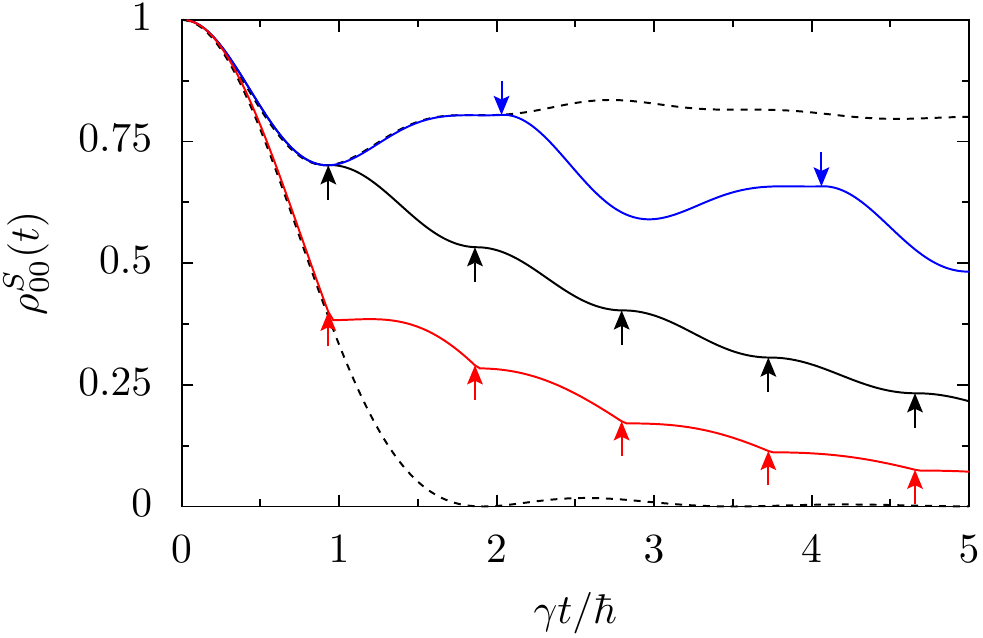}
 \caption{The time dependence of the probability to find the particle on site~0
   is shown for a chain consisting of 15 sites. The lower dashed curve and the
   lowest (red) solid curve refer to $\epsilon/\gamma=0$ while the upper dashed
   curve and the two upper (black and blue) solid curves have been obtained for
   an energy shift of site~0 of $\epsilon/\gamma=\pi$. The solid curves show the
   time evolution for measurements with $\gamma t_\text{m}/\hbar=\pi/100$ or
   $g/\gamma=100$ at times indicated by the arrows. For the two lower (red and black)
   solid curves, the periods of free dynamics are of length $\gamma t_\text{f}/\hbar
   = 0.9$ while for the upper (blue) solid curve, we have chosen
   $\gamma t_\text{f}/\hbar=2$. The dashed curves result from the free dynamics
   on the chain without any measurements. For all curves we have set $\delta=0$.}
 \label{fig:repmeas}
\end{figure}

First, in figure~\ref{fig:repmeas}, we compare the time dependence of
$\rho_{00}^\text{S}$ for different energy offsets $\epsilon$ on site~0 and
different durations $t_\text{f}$ of the free dynamics. The strength of the
coupling between system and apparatus is always given by $g/\gamma=100$
corresponding to a measurement time $\gamma t_\text{m}/\hbar=\pi/100$ and
$\delta=0$. The times at which measurements occur are indicated by arrows.

In the lower (red) solid curve, all on-site energies are degenerate and the
time between measurements is $\gamma t_\text{f}/\hbar=0.9$. Up to the first
measurement, $\rho_{00}^\text{S}(t)$ follows the free dynamics for
$\epsilon/\gamma=0$ depicted by the lower dashed curve. Comparing the two curves
one sees that the measurements tend to hinder the decay of the population on
site~0.

It is interesting to compare the case of degenerate on-site energies with the
case where the site~0 is shifted in energy by an amount $\epsilon/\gamma=\pi$.
The corresponding free evolution of $\rho_{00}^\text{S}(t)$ is given by the
upper dashed curve. Clearly, the energy shift makes it difficult for the
particle to leave its initial site. However, frequent measurements like in the
middle (black) solid curve for $\gamma t_\text{f}/\hbar=0.9$ assist the decay
of the population on site~0. Increasing the intervals between the measurements,
the population decay slows down as can be seen from the upper (blue) solid
curve which has been obtained for $\gamma t_\text{f}/\hbar=2$.

\begin{figure}
 \includegraphics[width=\columnwidth]{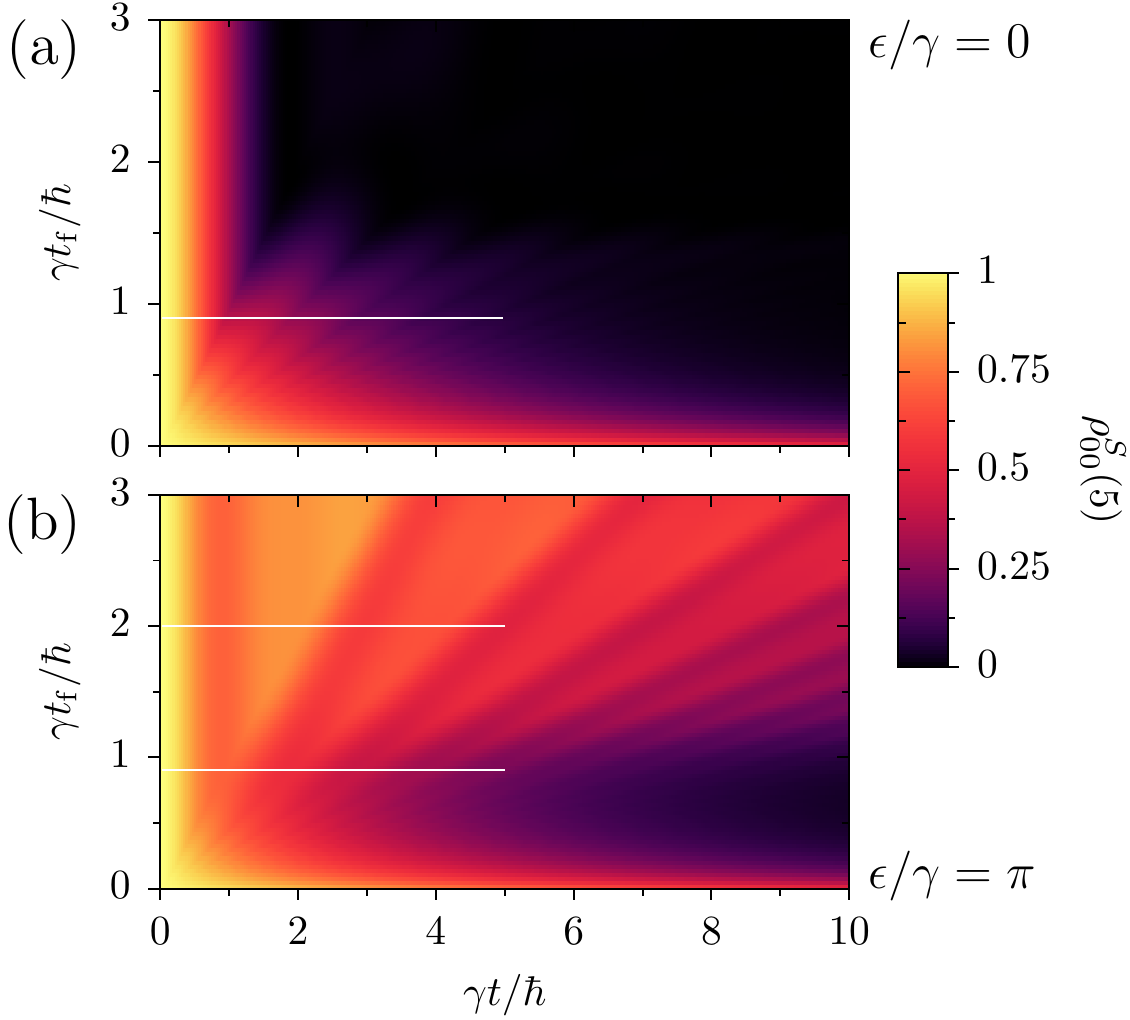}
 \caption{The occupation of site~0 is shown as function of time $t$ and the
   period of free evolution $t_\text{f}$ for finite-time measurements of a
   short duration $\gamma t_\text{m}/\hbar=\pi/100$. The chain consists of 15
   sites with the energy offset (a) $\epsilon/\gamma=0$ and (b) $\epsilon/
   \gamma=\pi$. In the coupling Hamiltonian we have set $\delta=0$. The white
   lines indicate the cuts displayed as solid curves in figure~\ref{fig:repmeas}.}
 \label{fig:zenomap}
\end{figure}

A more complete picture of the time evolution of the population on site~0 for
fixed measurement time $\gamma t_\text{m}/\hbar=\pi/100$ or $g/\gamma=100$ and
variable period of free dynamics of length $\gamma t_\text{f}/\hbar$ is given
in figure~\ref{fig:zenomap}(a) for $\epsilon/\gamma = 0$ and in
figure~\ref{fig:zenomap}(b) for $\epsilon/\gamma=\pi$. In both figures, the
parameter $\delta$ is set to zero but, because of the short measurement time,
its value does not affect the results much. The horizontal white lines indicate
the cuts represented by the solid curves in figure~\ref{fig:repmeas}.

If all on-site energies are degenerate, the particle can easily leave its
initial site~0 and move along the chain. The corresponding fast decay of
$\rho_{00}^\text{S}$ can clearly be seen in figure~\ref{fig:zenomap}(a) for
sufficiently large values of $t_\text{f}$. However, as the distance between
the measurements is decreased, the decay of $\rho_{00}^\text{S}$ is slowed
down. For very small values of $t_\text{f}$, the particle remains on site~0
for a very long time thereby manifesting the quantum Zeno effect.

The scenario just described should be contrasted with the case where the energy
of site~0 is shifted with respect to the energies of the other sites of the
chain. Comparing figure~\ref{fig:zenomap}(b), where $\epsilon/\gamma=\pi$, with
the previously discussed figure~\ref{fig:zenomap}(a) we note, that they
resemble each other for small values of $t_\text{f}$. With increasing
$t_\text{f}$, the decay of the population on site~0 accelerates and the quantum
Zeno effect becomes less effective. Increasing $t_\text{f}$ even further, the
free dynamics becomes dominant and the nonzero value of the energy shift
$\epsilon$ tends to suppress the decay of $\rho_{00}^\text{S}$. As a consequence,
the decay is slowest for $\gamma t_\text{f}/\hbar\lesssim 1$, i.e.\ when the
time between measurements is somewhat smaller than the time scale of the free
dynamics.

\begin{figure}
 \includegraphics[width=\columnwidth]{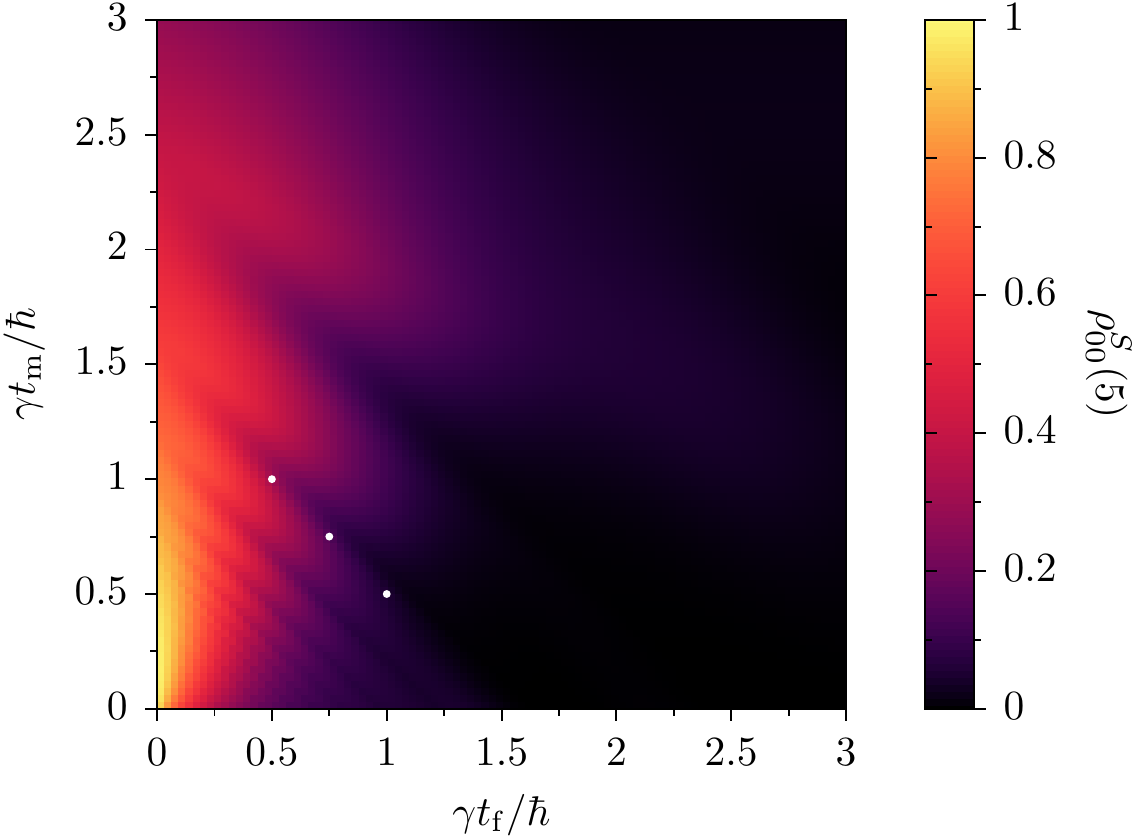}
 \caption{The occupation of site~0 at time $\gamma t/\hbar=5$ is shown as a
   function of the measurement duration $t_\text{m}$ and the period of free
   evolution $t_\text{f}$ for a chain consisting of 15 sites. All on-site
   energies are degenerate, $\epsilon/\gamma=0$, and $\delta=3/2$. The white
   points mark the parameters for which the time evolution of
   $\rho_{00}^\text{S}$ is shown in figure~\ref{fig:repfintime}.} 
 \label{fig:tm_tf}
\end{figure}

\begin{figure}[t]
 \includegraphics[width=\columnwidth]{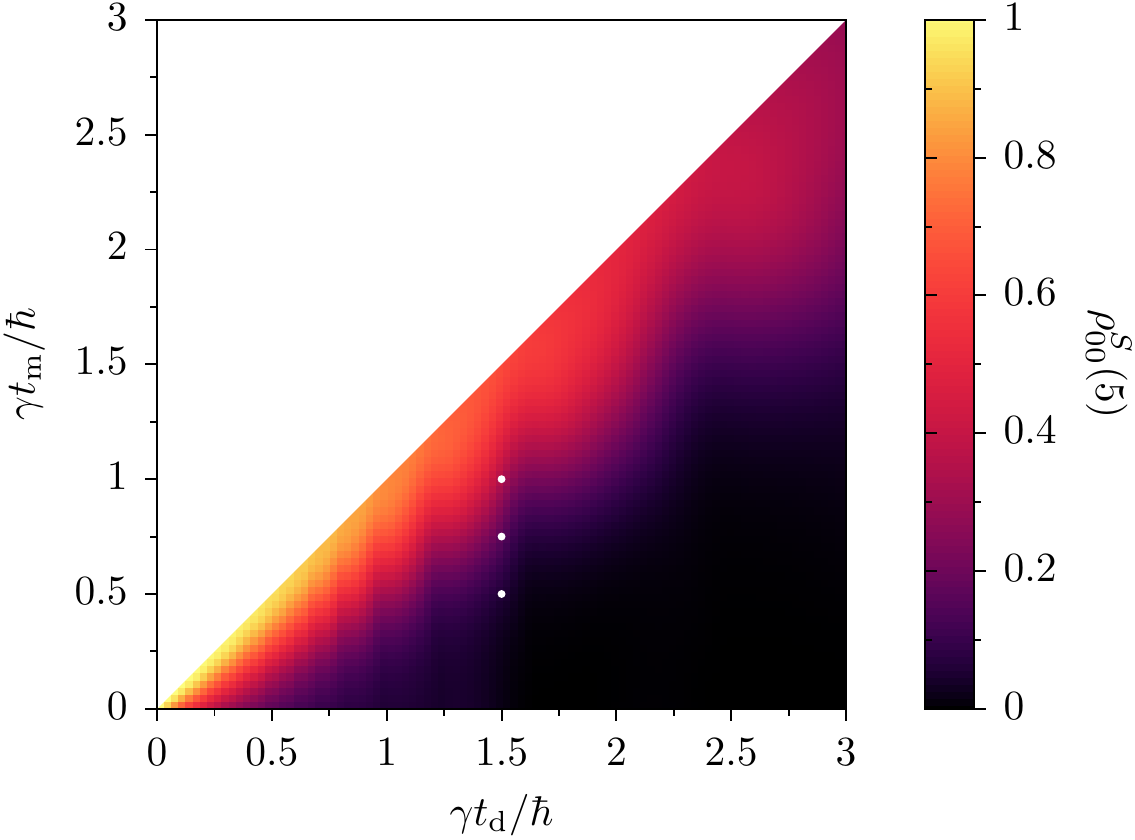}
 \caption{The data from figure~\ref{fig:tm_tf} are now shown as a function
   of the measurement duration $t_\text{m}$ and the time
   $t_\text{d}=t_\text{f}+t_\text{m}$ between the beginning of two subsequent
   measurements. In the upper left white triangle, $t_\text{f}$ would be
   negative. The white points mark the parameters for which the time evolution
   of $\rho_{00}^\text{S}$ is shown in figure~\ref{fig:repfintime}.} 
 \label{fig:tm_td}
\end{figure}

So far, we have kept the measurement time $t_\text{m}$ constant. Now, we want
to study the decay of $\rho_{00}^\text{S}$ as a function of the measurement
time and the time between measurements. Since the time between measurements can
be specified either in terms of $t_\text{f}$ or $t_\text{d}$ (cf.
figure~\ref{fig:timescales}), we obtain two different representations shown in
figures~\ref{fig:tm_tf} and \ref{fig:tm_td}. In these figures, the decay of
$\rho_{00}^\text{S}$, relative to its initial value $\rho_{00}^\text{S}(0)=1$,
is represented by its value at time $\gamma t/\hbar=5$. Of course, only for
special values of $t_\text{m}$ and $t_\text{f}$ will the time $\gamma
t/\hbar=5$ coincide with the beginning or end of a measurement process. We
consider a setup where all on-site energies are degenerate, i.e.\
$\epsilon/\gamma=0$. Furthermore, we choose $\delta=3/2$ which results in a
strong suppression of the decay of $\rho_{00}^\text{S}$ during the measurement
process.

In figure~\ref{fig:tm_tf}, the occupation on site~0 is represented as a
function of $t_\text{m}$ and $t_\text{f}$. For short times $t_\text{f}$ between
measurements and measurement durations of up to $\gamma t_\text{m}/\hbar\sim
1$, one finds a strong inhibition of the decay of the occupation. This decay
weakens as the period of free system evolution is increased.  For long
measurement durations $t_\text{m}$, in view of (\ref{eq:measurement_time_2})
the coupling between the system and the apparatus is very weak. The time
dependence of the occupation on site~0 is then dominated by the system dynamics.
The occupation is thus strongly suppressed at $\gamma t/\hbar=5$.

\begin{figure}
 \includegraphics[width=\columnwidth]{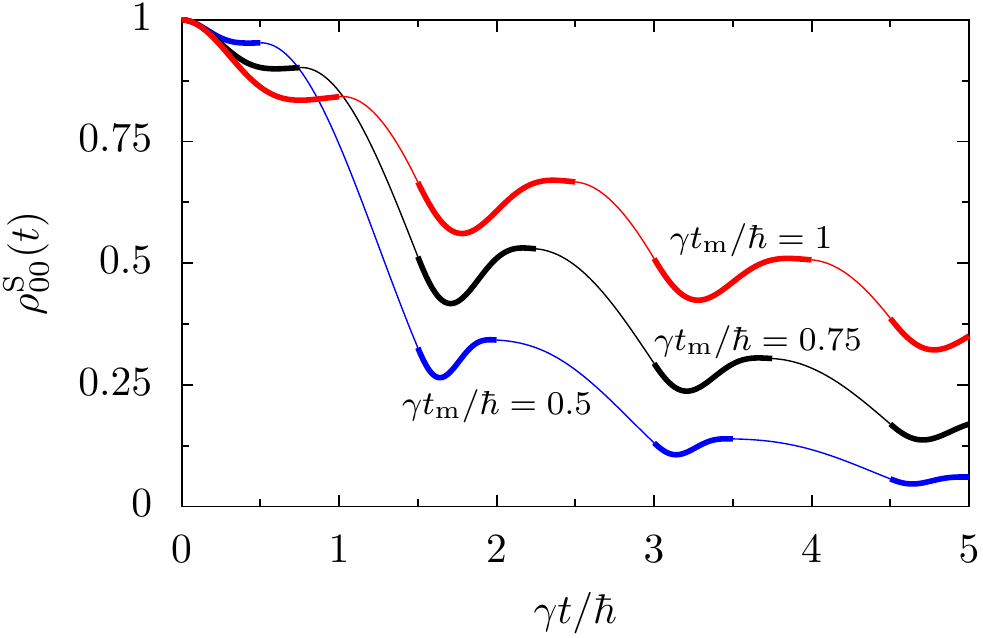}
 \caption{The occupation of site~0 of a chain consisting of 15 sites with
   $\epsilon/\gamma=0$ is shown as a function of time for measurement durations
   $\gamma t_\text{m}/\hbar=0.5, 0.75,$ and $1$. The time evolution during
   the measurement processes is depicted by the thick line segments. The time
   between the beginning of subsequent measurements is $\gamma t_\text{d}/\hbar=1.5$
   for all three curves and $\delta=\frac{3}{2}$.}
 \label{fig:repfintime}
\end{figure}

Interestingly, for sufficiently large values of $\delta$ like the one chosen
for figures~~\ref{fig:tm_tf}--\ref{fig:repfintime}, there exists an
intermediate regime for $t_\text{m}$ where $\rho_{00}^\text{S}$ as a function
of $t_\text{f}$ does not decrease as fast as in the limits of small and large
$t_\text{m}$. In this regime, the energetic degeneracy of site~0 and the other
sites is lifted sufficiently strongly for a long enough time to lead to an
appreciable suppression of the decay.

This effect is also visible in figure~\ref{fig:tm_td} where the same data are
represented as in figure~\ref{fig:tm_tf} but now as a function of $t_\text{m}$
and $t_\text{d}$. Since the time between the beginning of subsequent
measurement processes has to be larger than the measurement time itself, only
the lower right triangle contains data. The quantum Zeno effect is visible for
small values of $t_\text{d}$ close to the diagonal. For not too large fixed
times $t_\text{d}$ between the beginning of subsequent measurements, an
increase of the measurement time $t_\text{m}$ leads to an increase of
$\rho_{00}^\text{S}$ at a fixed time $t$. For large values of $t_\text{d}$, 
the reflection of the particle at the far end of the chain will become relevant.

Figure~\ref{fig:repfintime} shows how the values of $\rho_{00}^\text{S}(5)$
depicted in figures~\ref{fig:tm_tf} and \ref{fig:tm_td} are approached as a
function of time for the parameters indicated in the latter figures by white
points. The time between subsequent measurements $\gamma t_\text{d}/\hbar=1.5$
is kept fixed while the measurement time is different for the three curves. The
time evolution during the measurement periods is indicated by thick lines.
Except for short times, an increase of the measurement time will help to
inhibit the decay of the population on site~0. As already indicated above, this
dependence on the measurement time will only be noticeable if the energy shift
induced by the coupling to the measurement apparatus is sufficiently strong,
i.e.\ if $\delta$ is sufficiently large.

\section{Conclusions}
\label{sec:conclusions}

We have studied the decay of the population on a lattice site of a
one-dimensional chain under the influence of a local finite-time measurement.
Our setup provides a simple model to explore the interplay between the system
dynamics and the coupling to the measurement apparatus. For the special case of
a two-site chain, we have shown analytically, that the limit of infinite
coupling to the apparatus leads to a projective measurement. For finite values
of the coupling constant, we found that a shift $\delta$ of the eigenspectrum
of the observable conjugate to the pointer variable modifies the on-site energy
during the measurement in a significant way. Depending on the bare on-site
energy and the value of $\delta$, the decay of the population on the measured
site can be hindered or facilitated. If the measured site is not energetically
degenerate with the other sites, numerical results for a longer chain showed
that the decay is fastest for an intermediate value of the time between
measurements. While for shorter periods of free dynamics, the quantum Zeno
effect dominates, it is the unitary dynamics of the system which suppresses the
decay if measurements occur infrequently. On the other hand, for a chain with
degenerate energies, the coupling-induced shift of the on-site energy leads to
a suppression of the decay for intermediate values of the measurement time.  In
contrast, for short measurement times, the shift is effective only during a too
short time while for large measurement times, the coupling to the apparatus is
weak and the shift becomes again irrelevant.

\end{document}